\documentclass[%
preprint,
 amsmath,amssymb,
aps,
]{revtex4-1}

\usepackage{graphicx}
\usepackage{dcolumn}
\usepackage{bm}
\usepackage[breaklinks=true,colorlinks=true,linkcolor=blue,urlcolor=blue,citecolor=blue]{hyperref}
\usepackage{color}

\begin{document}
\preprint{arXiv-MET-V1}

\title{An analytical algorithm for 3D magnetic field mapping of a watt balance magnet}
\author{Zhuang Fu$^{1,2}$,
Zhonghua Zhang$^{2}$,
Zhengkun Li$^{2}$,
Wei Zhao$^{1}$,
Bing Han$^{2}$,
Yunfeng Lu$^{2}$,
Shisong Li$^{1}$}
 \altaffiliation[]{1. Department of Electrical Engineering, Tsinghua University, Beijing 100084, China\\
 2. National Institute of Metrology, Beijing 100029, China}
 \email{leeshisong@sina.com}

\date{\today}

\begin{abstract}
A yoke-based permanent magnet, which has been employed in many watt balances at national metrology institutes, is supposed to generate strong and uniform magnetic field in an air gap in the radial direction. However, in reality the fringe effect due to the finite height of the air gap will introduce an undesired vertical magnetic component to the air gap, which should either be measured or modeled towards some optimizations of the watt balance. A recent publication, i.e., {\it Metrologia} 52(4) 445 \cite{lishisong}, presented a full field mapping method, which in theory will supply useful information for profile characterization and misalignment analysis. This article is an additional material of \cite{lishisong}, which develops a different analytical algorithm to represent the 3D magnetic field of a watt balance magnet based on only one measurement for the radial magnetic flux density along the vertical direction, $B_r(z)$. The new algorithm is based on the electromagnetic nature of the magnet, which has a much better accuracy.
\end{abstract}

\maketitle

\section{Introduction}
\label{sec1}
The watt balance, an experiment proposed by Dr B. P. Kibble in 1976 \cite{Kibble1976}, is widely employed at National Metrology Institutes (NMIs) for precisely measuring the Planck constant $h$ towards the redefinition of one of the SI base units, the kilogram \cite{ShisongLi2012}.
The new definition of the kilogram will be realized by fixing the numerical value of the Planck constant \cite{new_Planck}, which is expected to be determined with a relative uncertainty of two parts in $10^8$. The role of a watt balance experiment is to transfer the mass standard from the only mass in Bureau International des Poids et Mesures (BIPM), i.e., the international prototype of kilogram (IPK), to a value that makes the Planck constant exactly equal to $6.62606...\times 10^{-34}$ J$\cdot$s.

The detailed origin, principle and recent progress of the watt balance is presented in several review papers, e.g., \cite{Li12, Stainer13, Stock13}. Here a brief summary of the measurement is given. The watt balance is operated in two separated measurement modes, conventionally named as the weighing mode and the velocity mode. In the weighing mode, the magnetic force produced by a coil with DC current in the magnetic field is balanced by the gravity of a test mass, and a force balance equation can be written as
\begin{equation}
BLI=mg,
\label{1}
\end{equation}
where $B$ denotes the magnetic flux density, $I$ the current in the coil, $L$ the wire length of the coil, $m$ the test mass and $g$ the local gravitational acceleration. In the velocity mode, the coil moves along the vertical direction in the same magnetic field, and generates an induced voltage, i.e.
\begin{equation}
BL\upsilon=\varepsilon,
\label{2}
\end{equation}
where $\upsilon$ is the coil velocity and $\varepsilon$ the induced voltage. By combining equations (\ref{1}) and (\ref{2}), the geometrical factor $BL$ can be eliminated and a virtual watt balance equation is obtained as
\begin{equation}
\varepsilon I=mg\upsilon.
\label{3}
\end{equation}
In equation (\ref{3}), the induced voltage $\varepsilon$ is measured by a Josephson voltage standard (JVS) linked to the Josephson effect, i.e.
\begin{equation}
\varepsilon=\frac{f_1 h}{2e},
\label{4}
\end{equation}
where $f_1$ denotes a known frequency, $e$ the electron charge. The current $I$ is measured by the Josephson effect in conjunction with the quantum Hall effect as
\begin{equation}
I=\frac{U}{I}=\left(\frac{f_2 h}{2e}\right)\left(\frac{h}{n_0e^2}\right)^{-1}=\frac{f_2n_0e}{2},
\label{5}
\end{equation}
where $U$ is the voltage drop on a resistor $R$ in series with the coil, $f_2$ the known frequency, and $n_0$ an integer number. A combination of equations (\ref{3})-(\ref{5}) yields the expression of the Planck constant $h$ in the SI unit as
\begin{equation}
h=\frac{4mg\upsilon}{f_1f_2n}.
\label{6}
\end{equation}
It has been known that all the quantities on the right side of (\ref{6}) can be measured with a relative uncertainty lower than one part in $10^{8}$, and therefore, on the current stage, the Planck constant $h$ is expected to be determined by watt balances with a relative uncertainty less than $2\times10^{-8}$ for achieving the purpose of redefining the kilogram.

\begin{figure}
\center
\includegraphics[width=0.45\textwidth]{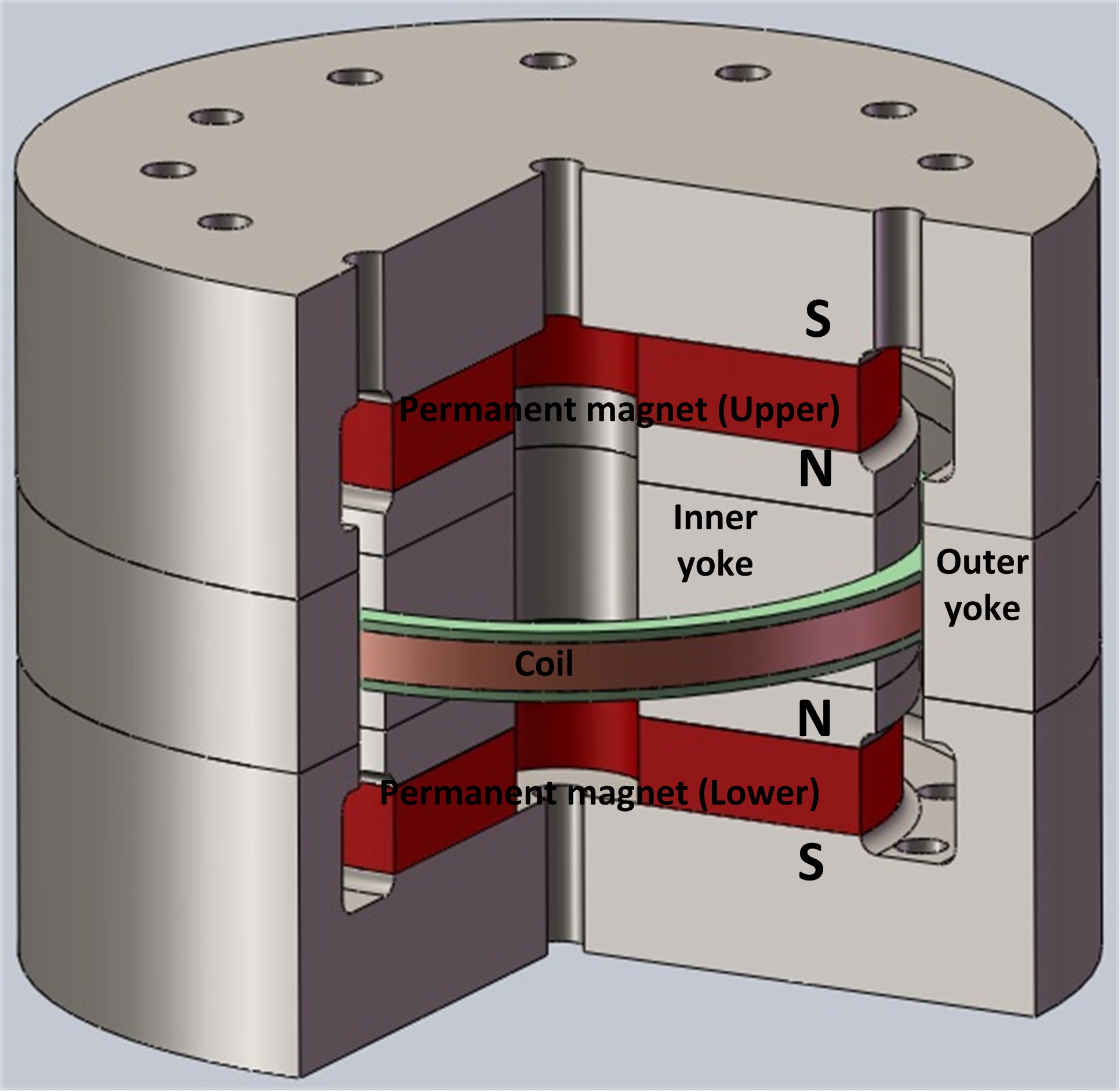}
\caption{A typical magnet construction in watt balances. The magnetic field is created by two permanent magnets whose N poles face toward each other. The coil is operated in the narrow air gap region where the majority of the magnetic flux density is in radial direction.}
\label{fig2}
\end{figure}

The watt balance is considered as one of the most successful experiments to precisely determine the Planck constant $h$ and is employed in many NMIs, e.g. \cite{NPL, NIST, METAS, BIPM, LNE, NIM, MSL, NRC, KRISS}.
In order to generate a kilogram level magnetic force to balance the gravity of a mass while keep the power assumption of the coil as low as possible in the weighing mode, a strong magnetic field, e.g. 0.5T, is required at the coil position for the watt balance. Accordingly, permanent magnets with high permeability yokes are introduced to achieve the strong magnetic field. One of such magnetic systems (shown in figure \ref{fig2}), developed by the BIPM watt balance group \cite{Stock2006}, is the most preferred. In the shown magnet construction, two permanent magnet rings with opposite magnetization poles are installed inside the inner yoke. The magnetic flux of the permanent magnet rings is guided by high permeability yokes through the air gap. As the work area of the coil, the air gap, is designed to be long and narrow, a strong magnetic field with good uniformity can be generated along the radial direction $r$. As the total flux through the air gap is roughly a constant, the magnetic flux density along the radial direction, $B_r(r)$, decays along $r$ direction following an approximative $1/r$ relation. In a $1/r$ magnetic field, it can be proved that the $BL$ of the coil is a constant, independent to any coil deformation or horizontal displacements \cite{lishisongcoil}. As the magnetic flux is closed with yokes, the shown watt balance magnet has a good feature of self shielding, lowing additional flux exchange between inside and outside of the magnet. Based on these advantages, the shown magnet has been widely adopted by other NMIs, such as the Federal Institute of Metrology (METAS), Switzerland \cite{Baumann2013}, the National Institute of Standards and Technology (NIST), USA \cite{Seifert2014}, the Measurement Standards Laboratory (MSL), New Zealand \cite{Sutton2014} and the Korea Research Institute of Standards and Science (KRISS), South Korea \cite{KRISS}.

In reality, since the air gap length is finite and hence the $1/r$ decay of the magnetic field $B_r(r)$ is not true over the whole vertical range, i.e., the fringe effect will introduce a vertical magnetic component $B_z$ at any coil positions where $z\neq0$. This fact indicates that the unknown magnetic field component can cause alignment problems \cite{Robinson}, e.g., the vertical magnetic field component can exert undesired radial force or torque in the weighing mode and could produce additional voltage in the velocity mode. Considering a strict requirement of measurement accuracy in watt balances, any aspect may bring a systematic error should be carefully analyzed. An accurate 3D mapping of the magnetic field in the air gap is a useful tool to provide important information on systematic error elimination.
At least three benefits will be created by a full air gap 3D magnetic field mapping. The first is the misalignment error relaxation: the value of geometrical factor $BL$ in equations (\ref{1}) and (\ref{2}) can be actually affected by parasitic coil motions when the fringe field is considered. This misalignment error can be corrected if the 3D magnetic field profile is obtained, which in theory can relax the alignment requirement of a watt balance. The second benefit is that with knowing a full field profile, a best coil diameter can be chosen in order to employ the maximum flat $B_r(z)$ profile in the air gap, and a flat $B_r(z)$ profile can reduce uncertainties in both velocity and voltage measurements. Thirdly, the damping of the coil is usually applied in the breath between weighing and velocity sweeps. During the breath, the damping device is either at the top or the bottom of the air gap, where a coupling between the damping current and both magnetic components (the radial magnetic field $B_r$ and vertical magnetic field $B_z$) should be considered. In this case, the 3D magnetic field can supply better feedback to the current parameters for simultaneously damping all dimensional motions of the coil.

As is mentioned in \cite{lishisong}, the narrow air gap makes it difficult to directly measure the global magnetic field profiles. Only profiles of the radial magnetic flux density along the vertical direction, $B_r(z)$, can be precisely measured by either a high resolution magnetic probe or the gradient coil (GC) method \cite{Seifert2014}. A polynomial estimation algorithm based on at least two measurements of $B_r(z)$ profiles has been developed in \cite{lishisong} to calculate the global magnetic field in the air gap. The algorithm essentially approximates the magnetic field lines by a polynomial fitting and the approximation precision depends on the fitting order, i.e. the quantities of the measured $B_r(z)$. A higher accuracy requires more $B_r(z)$ profile measurements, which in reality is difficult to be done at different radii of the narrow air gap. Besides, the algorithm presented in \cite{lishisong} would be much more complicated when the estimator order $N$ increases, e.g., $N>2$. We later noticed the information in measured $B_r(z)$ is not fully utilized in the polynomial estimation algorithm, and hence in theory it is possible to develop an analytical algorithm based on only one $B_r(z)$ measurement. This motivation leads to an improved analytical algorithm to map the 3D magnetic field in the air gap region, which has been presented in this article. The new algorithm is based on fundamental electromagnetic natures of the magnet and has advantages in both convenience and accuracy.

The rest of this paper is organized as following. In section \ref{sec2}, the new analytical algorithm is presented. In section \ref{sec3}, numerical simulations are employed to verify the accuracy of the analytical algorithm. Some discussions on the watt balance magnet design are shown in section \ref{sec4} and a conclusion is drawn in section \ref{sec5}.

\section{Analytical algorithm}
\label{sec2}
\begin{figure}
\center
\includegraphics[width=0.4\textwidth]{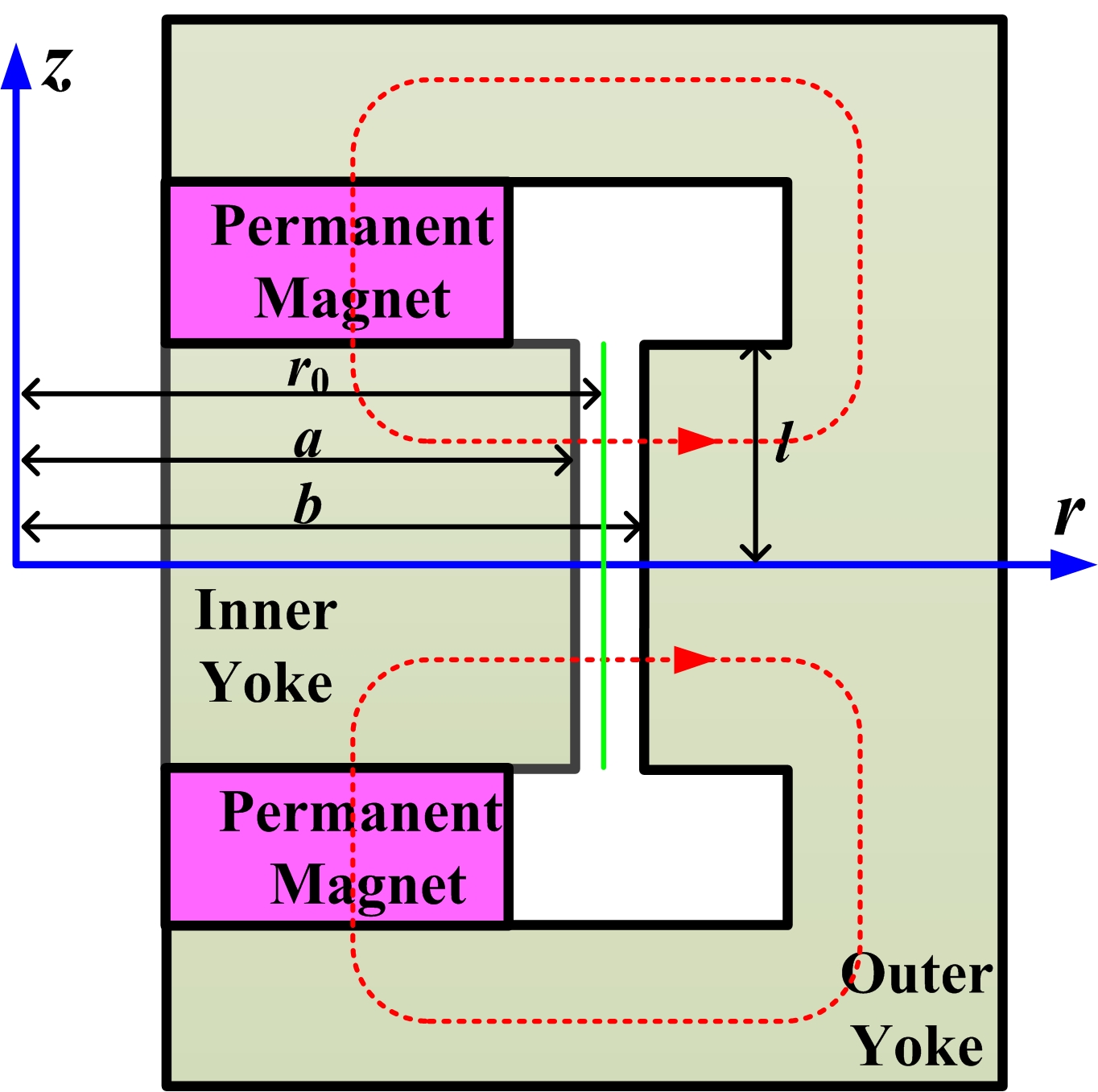}
\caption{Sectional view of the air gap structure used in watt balances. The magnetic field (the red dotted line) is generated by two permanent magnets whose N poles face toward each other. $B_r(r_0, z)$ is measured at the horizontal coordinate $r_0$ (the green line).}
\label{fig3}
\end{figure}

A detailed dimension of the air gap in the watt balance magnet is shown in figure \ref{fig3}. The area $r\in(a, b)$, $z\in(-l, l)$ is our model region where $a$  and $b$ is the inner and outer radii of the air gap and $l$ is a half of the air gap height. The radial magnetic flux density along the vertical direction $B_r(r_0,z)$, or the radial magnetic field along the vertical direction $H_r(r_0,z)$, can be measured by a GC or a magnetic probe at the radial coordinate $r_0$.

In the analysis, the magnetic scalar potential $\varphi_m$ is chosen as a quantity to be solved. The magnetic field $\overrightarrow{H}(r,z)$ can be expressed as the negative gradient of the magnetic scalar potential, i.e., $\overrightarrow{H}(r,z)=-\nabla \varphi_m$. $\overrightarrow{H}(r,z)$ is the sum of two components in an axisymmetrical coordinate system as
\begin{eqnarray}
\begin{array}{l}
\overrightarrow{H}(r,z)
=H_r(r,z)\overrightarrow{e_r}+H_z(r,z)\overrightarrow{e_z}
\label{eq1}
\end{array}
\end{eqnarray}
where $\overrightarrow{e_r}$ and $\overrightarrow{e_z}$ denote the unit vectors in the $r$ and $z$ directions; $H_r(r,z)$ and $H_z(r,z)$ are two components of the magnetic field in $r$ and $z$ directions. As $\overrightarrow{H}(r,z)=-\nabla \varphi_m$, equation (\ref{eq1}) can also be written as
\begin{eqnarray}
\begin{array}{l}
\overrightarrow{H}(r,z)
=\displaystyle-\frac{\partial \varphi_m}{\partial r}\overrightarrow{e_r}-\frac{\partial \varphi_m}{\partial z}\overrightarrow{e_z}.
\label{eq1.1}
\end{array}
\end{eqnarray}
The magnetic scalar potential $\varphi_m$ in the air gap area, i.e. $r\in$ ($a$, $b$), $z\in$ ($-l$, $l$), can be described by the Laplace's equation as
\begin{equation}
\nabla^2\varphi_m=\frac{1}{r}\frac{\partial}{\partial r}\left(r\frac{\partial \varphi_m}{\partial r}\right)+\frac{\partial^2\varphi_m}{\partial z^2}=0.
\label{eq2}
\end{equation}
The separation of variables is applied to solve equation (\ref{eq2}). In this case, the magnetic scalar potential $\varphi_m$ is supposed to be expressed as the product of two independent functions $R(r)$ and $Z(z)$, where $R(r)$ and $Z(z)$ are functions of a single variable, i.e. $r$ and $z$. Substituting $\varphi_m=R(r)Z(z)$ into equation (\ref{eq2}) yields
\begin{equation}
\frac{1}{R}\frac{1}{r}\frac{d}{{dr}}\left( {r\frac{{dR}}{{dr}}} \right) + \frac{1}{Z}\frac{{{d^2}Z}}{{d{z^2}}} = 0.
\label{sep_vars}
\end{equation}

Since the two terms on the left side of equation (\ref{sep_vars}) are independent in dimension, equation (\ref{sep_vars}) can be written as
\begin{equation}
\left\{ \begin{array}{l}
\displaystyle\frac{{{d^2}Z}}{{d{z^2}}} - {\kappa}Z = 0\\
\displaystyle\frac{1}{r}\frac{d}{{dr}}\left( {r\frac{{dR}}{{dr}}} \right) + {\kappa}R = 0,
\end{array} \right.
\end{equation}
where $\kappa$ is a constant. Note that the solution form of equation (\ref{sep_vars}) will be determined by the sign of $\kappa$. When $\kappa=0$, the solution can be written as
\begin{equation}
{\varphi _{m1}} = \left( {Az + B} \right)\left( {C\ln r + D} \right),
\end{equation}
where $A$, $B$, $C$ and $D$ are constants. When $\kappa>0$ (e.g. $\kappa=\lambda^2$), the solution can be obtained as
\begin{equation}
\begin{array}{l}
{\varphi _{m2}} = \left[ {{A_1}\cosh (\lambda z) + {B_1}\sinh (\lambda z)} \right]
\left[{{C_1}{J_0}(\lambda r) + {D_1}{Y_0}(\lambda r)} \right],
\end{array}
\end{equation}
where $\cosh ({\lambda}z)$ and $\sinh ({\lambda}z)$ denote hyperbolic cosine function and hyperbolic sine function; $J_0(\lambda r)$ and $Y_0(\lambda r)$ are the zeroth order Bessel functions of the first and second kinds, $A_{1}$, $B_1$, $C_1$, $D_1$ and $\lambda$ are all constants. When $\kappa<0$ (e.g. $\kappa=-\rho^2$), the solution can be written as
\begin{equation}
\begin{array}{l}
\varphi _{m3} = \left[ {{A_2}\cos (\rho z) + {B_2}\sin (\rho z)} \right]
\left[ {{C_2}{I_0}(\rho r) + {D_2}{K_0}(\rho r)} \right],
\end{array}
\end{equation}
where $I_0(\rho r)$ and $K_0(\rho r)$ denote the zeroth order modified Bessel functions of the first and second kinds, $A_{2}$, $B_{2}$, $C_{2}$, $D_{2}$ and $\rho$ are all constants.

The general solution of equation (\ref{sep_vars}) is the linear combination of $\varphi_{m1}$, $\varphi_{m2}$ and $\varphi_{m3}$, i.e.
\begin{eqnarray}
\begin{array}{l}
\varphi_m=(Az+B)(C\ln r+D)\\
+\sum\limits_{n=1}^{\infty}\left[A_{1n}\cosh (\lambda_n z)+B_{1n}\sinh (\lambda_n z)\right]
\left[C_{1n}J_0(\lambda_n r)+D_{1n}Y_0(\lambda_n r)\right]\\
+\sum\limits_{n=1}^{\infty}\left[A_{2n}\cos (\rho_n z)+B_{2n}\sin (\rho_n z)\right]
\left[C_{2n}I_0(\rho_n r)+D_{2n}K_0(\rho_n r)\right].
\label{eq.general}
\end{array}
\end{eqnarray}
Based on equations (\ref{eq1.1}) and (\ref{eq.general}), the vertical magnetic field component $H_z(r,z)$ can be expressed as
\begin{eqnarray}
\begin{array}{l}
\displaystyle H_z(r,z)=-\frac{\partial \varphi_m}{\partial z}\\
=-A(C\ln r+D)\\
-\sum\limits_{n=1}^{\infty}\lambda_n\left[A_{1n}\cosh (\lambda_n z)+B_{1n}\sinh (\lambda_n z)\right]
\left[C_{1n}J_0(\lambda_n r)+D_{1n}Y_0(\lambda_n r)\right]\\
+\sum\limits_{n=1}^{\infty}\rho_n\left[A_{2n}\sin (\rho_n z)-B_{2n}\cos (\rho_n z)\right]
\left[C_{2n}I_0(\rho_n r)+D_{2n}K_0(\rho_n r)\right].
\label{eq4}
\end{array}
\end{eqnarray}

Since the yoke permeability in the presented watt balance magnet is very high, two yoke-air boundaries, i.e. $r=a$ and $r=b$, can be considered as equipotential surfaces. As a result, the vertical magnetic field along the vertical direction should be zero at both $r=a$ and $r=b$, i.e., $H_z(a,z)=H_z(b,z)=0$. Using this boundary condition, the following equations are established, i.e.
\begin{equation}
\left\{
\begin{array}{l}
A(C\ln a+D)=0  \\
A(C\ln b+D)=0,
\label{eq.condition11}
\end{array}
\right.
\end{equation}

\begin{equation}
\left\{
\begin{array}{l}
C_{1n}J_0(\lambda_n a)+D_{1n}Y_0(\lambda_n a)=0  \\
C_{1n}J_0(\lambda_n b)+D_{1n}Y_0(\lambda_n b)=0,
\label{eq.condition12}
\end{array}
\right.
\end{equation}
and
\begin{equation}
\left\{
\begin{array}{l}
C_{2n}I_0(\rho_n a)+D_{2n}K_0(\rho_n a)=0  \\
C_{2n}I_0(\rho_n b)+D_{2n}K_0(\rho_n b)=0.
\label{eq.condition13}
\end{array}
\right.
\end{equation}

In equation (\ref{eq.condition11}), since $C\ln r +D$ is a function of monotonicity where $C\neq0$, we have $C\ln a +D\neq C\ln b +D$ when $a<b$, and hence $A=0$. In equation (\ref{eq.condition12}), we can set $C_{1n}=Y_0(\lambda_n b)$ and $D_{1n}=-J_0(\lambda_n b)$ to ensure the condition, i.e., $C_{1n}J_0(\lambda_n b)+D_{1n}Y_0(\lambda_n b)=0$, to be always satisfied. In this case, $\lambda_n$ should be set to values to establish another condition, i.e. $J_0(\lambda_n a)Y_0(\lambda_n b)-J_0(\lambda_n b)Y_0(\lambda_n a)=0$. In equation (\ref{eq.condition13}), because $I_0(\rho_n r)$ is a monotone increasing function while $K_0(\rho_n r)$ is a monotone decreasing function, we have $I_0(\rho_n a)/I_0(\rho_n b)<1$ and $K_0(\rho_n a)/K_0(\rho_n b)>1$. Therefore, $I_0(\rho_n a)K_0(\rho_n b)-I_0(\rho_n b)K_0(\rho_n a)\neq0$ and $C_{2n}=D_{2n}=0$ is obtained.

According to equations (\ref{eq.condition11})-(\ref{eq.condition13}), equation (\ref{eq.general}) can be simplified as
\begin{eqnarray}
\begin{array}{l}
\varphi_m=(BC\ln r+BD)\\
+\sum\limits_{n=1}^{\infty}\left[A_{1n}\cosh (\lambda_n z)+B_{1n}\sinh (\lambda_n z)\right]
\left[J_0(\lambda_n r)Y_0(\lambda_n b)-J_0(\lambda_n b)Y_0(\lambda_n r)\right].~~~~
\label{eq.after_condition1}
\end{array}
\end{eqnarray}
It can be seen in equation (\ref{eq.after_condition1}) that $B$ is not independent from $C$ and $D$. Without loss of generality, $B$ can be set as 1 and equation (\ref{eq.after_condition1}) can be rewritten as
\begin{eqnarray}
\begin{array}{l}
\varphi_m=C\ln r+D\\
+\sum\limits_{n=1}^{\infty}\left[A_{1n}\cosh (\lambda_n z)+B_{1n}\sinh (\lambda_n z)\right]
\left[J_0(\lambda_n r)Y_0(\lambda_n b)-J_0(\lambda_n b)Y_0(\lambda_n r)\right].~~~~
\label{eq.after_condition11}
\end{array}
\end{eqnarray}

Making the magnetic potential at $r=b$ be equal to zero, i.e. $\varphi_m(b, z)=0$, equation (\ref{eq.after_condition11}) can be written as
\begin{eqnarray}
\begin{array}{l}
\varphi_m=C(\ln r-\ln b)\\
+\sum\limits_{n=1}^{\infty}\left[A_{1n}\cosh (\lambda_n z)+B_{1n}\sinh (\lambda_n z)\right]
\left[J_0(\lambda_n r)Y_0(\lambda_n b)-J_0(\lambda_n b)Y_0(\lambda_n r)\right].~~~~
\label{eq.after_condition3}
\end{array}
\end{eqnarray}

The distribution of $\varphi_m$ is symmetrical about the line $z=0$ and thus the odd symmetrical function $\sinh (\lambda_n z)$ should be removed from the expression of $\varphi_m$. Then equation (\ref{eq.after_condition3}) is reduced to
\begin{equation}
\begin{array}{l}
\varphi_m=C(\ln r-\ln b)
+\sum\limits_{n=1}^{\infty}A_{1n}\cosh (\lambda_n z)
\left[J_0(\lambda_n r)Y_0(\lambda_n b)-J_0(\lambda_n b)Y_0(\lambda_n r)\right].
\label{eq.after_condition2}
\end{array}
\end{equation}

Using the measured profile $H_r(r_0, z)$, the remaining unknown constants in equation (\ref{eq.after_condition2}) can be solved. Based on equations (\ref{eq1}) and (\ref{eq.after_condition2}), $H_r(r_0, z)$ can be written as
\begin{eqnarray}
\begin{array}{l}
\displaystyle H_r(r_0,z)=\left.-\frac{\partial \varphi_m}{\partial r}\right|_{r=r_0}\\
\displaystyle=-\frac{C}{r_0}
+\sum\limits_{n=1}^{\infty}A_{1n}\lambda_n\cosh (\lambda_n z)
\left[J_1(\lambda_n r_0)Y_0(\lambda_n b)-J_0(\lambda_n b)Y_1(\lambda_n r_0)\right]\\
\displaystyle=C'+\sum\limits_{n=1}^{\infty}A'_n \cosh(\lambda_n z),
\label{eq.compare1}
\end{array}
\end{eqnarray}
where $J_1(\lambda_n r_0)$ and $Y_1(\lambda_n r_0)$ denote the first order Bessel functions of the first and second kinds; $C'=-C/r_0$; $A'_n=A_{1n}\lambda_n\left[J_1(\lambda_n r_0)Y_0(\lambda_n b)-J_0(\lambda_n b)Y_1(\lambda_n r_0)\right]$ ($n\geq1$). For convenience of expression, we can set $\lambda_0=0$ and $A'_0=C'$, and then $A'_0\cosh(\lambda_0 z)=C'$. As a result, equation (\ref{eq.compare1}) can be expressed as
\begin{eqnarray}
\begin{array}{l}
H_r(r_0,z)=\sum\limits_{n=0}^{\infty}A'_n \cosh(\lambda_n z).
\label{eq.compare11}
\end{array}
\end{eqnarray}

The values of $A'_n$ can be obtained by expanding $H_r(r_0,z)$ in forms of $\cosh(\lambda_n z)$. However, the implementation process is complicated because of the non-orthogonality of the base functions $\cosh(\lambda_n z)$, i.e.
\begin{equation}
\begin{array}{l}
\Big\langle\cosh(\lambda_m z),\cosh(\lambda_n z)\Big\rangle\\
\displaystyle=\int_{-h}^{h}\cosh(\lambda_m z)\cosh(\lambda_n z)dz\neq0~~(m \neq n),
\label{eq.after_condition85}
\end{array}
\end{equation}
where $\Big\langle f(z),g(z)\Big\rangle=\int_{-h}^{h}f(z)g(z)dz$ is defined as the inner product of two functions $f(z)$ and $g(z)$. To reduce the complexity due to the non-orthogonality, in this paper, the process of obtaining $A'_n$ is as follows: First, the base functions $\cosh(\lambda_n z)$ are transformed into orthogonal normalized base functions $e_n$, then $H_r(r_0,z)$ is expanded in forms of $e_n$ in a convenient way, and finally base functions $e_n$ are replaced with $\cosh(\lambda_n z)$. In this way, $H_r(r_0,z)$ can be easily expanded in forms of $\cosh(\lambda_n z)$ and hence $A'_n$ is obtained.

To practise the above idea, here the Gram-Schmidt Orthogonalization Procedure is employed to transform $\cosh(\lambda_n z)$ into orthogonal normalized base functions $e_n$. The definition of the norm of function $f(z)$ is expressed as
\begin{equation}
\|f(z)\|=\sqrt{\Big\langle f(z),f(z)\Big\rangle},
\end{equation}
then base functions $\cosh(\lambda_n z)$ are orthogonalized and normalized through a recursion formula, expressed as
\begin{equation}
\left\{
\begin{array}{l}
y_0=\cosh(\lambda_0 z)~~(n=0)  \\
e_0=y_0/\|y_0\|~~(n=0)\\
y_n=\cosh(\lambda_n z)-\sum\limits_{k=0}^{n-1}\Big\langle \cosh(\lambda_n z), e_k\Big\rangle e_k~~(n\geq1)\\
e_n=y_n/\|y_n\| ~~(n\geq1).
\label{eq.orthogonal_2}
\end{array}
\right.
\end{equation}
In recursion formula (\ref{eq.orthogonal_2}) the inner product $\big\langle e_m, e_n\big\rangle=0$ when $m\neq n$ and $\big\langle e_m, e_n\big\rangle=1$ when $m=n$.

After the orthogonalization and normalization procedure, $H_r(r_0,z)$ can be expanded in forms of $e_n$, i.e.
\begin{eqnarray}
\begin{array}{l}
H_r(r_0, z)=\sum\limits_{n=0}^{\infty}F_ne_n,
\label{eq.orthogonal_3}
\end{array}
\end{eqnarray}
where the coefficient $F_n$ can be calculated by solving the inner product of $H_r(r_0,z)$ and $e_n$, because
\begin{eqnarray}
\begin{array}{l}
F_n=\sum\limits_{m=0}^{\infty}F_m \Big\langle e_m,e_n\Big\rangle\\
=\Big\langle\sum\limits_{m=0}^{\infty}F_m e_m, e_n\Big\rangle\\
=\Big\langle H_r(r_0, z),e_n\Big\rangle.
\label{eq.orthogonal_31}
\end{array}
\end{eqnarray}

The last step of obtaining $A'_n$ is replacing $e_n$ in equation (\ref{eq.orthogonal_3}) with $\cosh(
\lambda_n z)$, and thus $e_n$ should be solved by expressing it with combinations of $\cosh(\lambda_n z)$. The relation between $\cosh(\lambda_n z)$ and $e_n$ has been shown in recursion formula (\ref{eq.orthogonal_2}). It is found difficult to directly express $e_n$ as a linear combination of $\cosh(\lambda_n z)$, but instead it is easy to write $\cosh(\lambda_n z)$ as a linear combination of $e_n$, i.e.
\begin{equation}
\left\{
\begin{array}{l}
\cosh(\lambda_0 z)=\|y_0\| e_0~~(n=0)  \\
\cosh(\lambda_n z)=\|y_n\| e_n+\sum\limits_{k=0}^{n-1}\Big\langle \cosh(\lambda_n z), e_k\Big\rangle e_k~~(n\geq1),
\label{eq.orthogonal_6}
\end{array}
\right.
\end{equation}
where $\|y_0\|$, $\|y_n\|$ and $\Big\langle \cosh(\lambda_n z), e_k\Big\rangle$ are constants that have been solved in equation (\ref{eq.orthogonal_2}).
Setting $\mathbf x$ and $\mathbf e$ as column vectors as
 \begin{equation}
 \mathbf x=\Big(\cosh(\lambda_0 z),\cosh(\lambda_1 z),...\Big)^T,
 \mathbf e=\Big(e_0,e_1,...\Big)^T,
 \end{equation}
 equation (\ref{eq.orthogonal_6}) can be written as the product of a matrix and the column vector $\mathbf{e}$, i.e.
\begin{equation}
\mathbf x=\mathbf M \mathbf e,
\label{matrix1}
\end{equation}
where $\mathbf M$ denotes a lower triangular matrix, i.e.
\begin{equation}
    \mathbf{M}=\left(
      \begin{array}{ccccc}
        \|y_0\| & 0 & 0 & 0 & \cdots \\
        \big\langle \cosh(\lambda_1 z),e_0\big\rangle  & \|y_1\| & 0 & 0 & \cdots \\
        \big\langle \cosh(\lambda_2 z),e_0\big\rangle  & \big\langle \cosh(\lambda_2 z),e_1\big\rangle & \|y_2\| & 0 & \cdots \\
        \vdots & \vdots & \vdots & \vdots & \ddots \\
        \vdots & \vdots & \vdots & \vdots & \ddots \\
      \end{array}
    \right).
    \label{eq.matrix}
\end{equation}
Based on equation (\ref{matrix1}), $e_n$ can be solved as the linear combination of $\cosh(\lambda_n z)$, i.e.
\begin{eqnarray}
\begin{array}{l}
e_n=\sum\limits_{i=0}^{\infty}\mathbf{M}^{-1}(n,i)\cosh(\lambda_i z),
\label{ex}
\end{array}
\end{eqnarray}
where $\mathbf{M}^{-1}(n,i)$ denotes the element on line numbered $n$ column numbered $i$ of $\mathbf{M}^{-1}$, the inverse of the matrix $\mathbf{M}$.
Knowing the $e_n$ expression with a linear combination of $\cosh(\lambda_n z)$, the unknown constants $A'_n$ in equation (\ref{eq.compare11}) can be solved. Replacing $e_n$ in equation (\ref{eq.orthogonal_3}) with equation (\ref{ex}), $H_r(r_0,z)$ is written as the combination of $\cosh(\lambda_n z)$, i.e.
\begin{eqnarray}
\begin{array}{l}
H_r(r_0, z)\\
=\sum\limits_{n=0}^{\infty}F_n\Big[\sum\limits_{i=0}^{\infty}\mathbf{M}^{-1}(n,i)\cosh(\lambda_i z)\Big]\\
=\sum\limits_{i=0}^{\infty}\Big[\sum\limits_{n=0}^{\infty}F_n\mathbf{M}^{-1}(n,i)\Big]\cosh(\lambda_i z)\\
=\sum\limits_{n=0}^{\infty}\Big[\sum\limits_{i=0}^{\infty}F_i\mathbf{M}^{-1}(i,n)\Big]\cosh(\lambda_n z),
\label{eq.compare2}
\end{array}
\end{eqnarray}
Comparing equations (\ref{eq.compare11}) and (\ref{eq.compare2}), $A'_n$ is solved as
\begin{eqnarray}
\begin{array}{l}
A'_n=\sum\limits_{i=0}^{\infty}F_i\mathbf{M}^{-1}(i,n).
\end{array}
\end{eqnarray}
Accordingly, the constant $C$ in equation (\ref{eq.compare1}) is calculated as
\begin{eqnarray}
\begin{array}{l}
C=-r_0 A'_0
=-r_0\left[\sum\limits_{i=0}^{\infty}F_i\mathbf{M}^{-1}(i,0)\right],
\label{eq.C}
\end{array}
\end{eqnarray}
and $A_{1n}$ in equation (\ref{eq.compare1}) is solved as
\begin{eqnarray}
\begin{array}{l}
\displaystyle A_{1n}=\frac{A'_n}{\lambda_n\left[J_1(\lambda_n r_0)Y_0(\lambda_n b)-J_0(\lambda_n b)Y_1(\lambda_n r_0)\right]}\\
\displaystyle=\frac{\sum\limits_{i=0}^{\infty}F_i\mathbf{M}^{-1}(i,n)}{\lambda_n\left[J_1(\lambda_n r_0)Y_0(\lambda_n b)-J_0(\lambda_n b)Y_1(\lambda_n r_0)\right]}.
\label{eq.A1n}
\end{array}
\end{eqnarray}
By a combination of equations (\ref{eq.after_condition2}), (\ref{eq.C}) and (\ref{eq.A1n}), the magnetic scalar potential $\varphi_m$ is obtained as
\begin{eqnarray}
\begin{array}{l}
\varphi_m=-r_0\left[\sum\limits_{i=0}^{\infty}F_i\mathbf{M}^{-1}(i,0)\right](\ln r-\ln b)\\
+\sum\limits_{n=1}^{\infty}\left[\sum\limits_{i=0}^{\infty}F_i\mathbf{M}^{-1}(i,n)\right]
\displaystyle\frac{\cosh (\lambda_n z)[J_0(\lambda_n r)Y_0(\lambda_n b)-J_0(\lambda_n b)Y_0(\lambda_n r)]}{\lambda_n\left[J_1(\lambda_n r_0)Y_0(\lambda_n b)-J_0(\lambda_n b)Y_1(\lambda_n r_0)\right]}.
\label{eq.solution}
\end{array}
\end{eqnarray}
Based on equation (\ref{eq1}), the two magnetic field components $H_z(r,z)$ and $H_r(r,z)$ can be calculated respectively as
\begin{equation}
\begin{array}{l}
\displaystyle H_z(r,z)=-\frac{\partial \varphi_m}{\partial z}\\
=-\sum\limits_{n=1}^{\infty}\left[\sum\limits_{i=0}^{\infty}F_i\mathbf{M}^{-1}(i,n)\right]
\displaystyle\frac{\sinh (\lambda_n z)\left[J_0(\lambda_n r)Y_0(\lambda_n b)-J_0(\lambda_n b)Y_0(\lambda_n r)\right]}{J_1(\lambda_n r_0)Y_0(\lambda_n b)-J_0(\lambda_n b)Y_1(\lambda_n r_0)},
\label{eq.Hz1}
\end{array}
\end{equation}
and
\begin{eqnarray}
\begin{array}{l}
\displaystyle H_r(r,z)=-\frac{\partial \varphi_m}{\partial r}
=\frac{r_0\sum\limits_{i=0}^{\infty}F_i\mathbf{M}^{-1}(i,0)}{r}\\
+\sum\limits_{n=1}^{\infty}\left[\sum\limits_{i=0}^{\infty}F_i\mathbf{M}^{-1}(i,n)\right]
\displaystyle\frac{\cosh (\lambda_n z)\left[J_1(\lambda_n r)Y_0(\lambda_n b)-J_0(\lambda_n b)Y_1(\lambda_n r)\right]}{J_1(\lambda_n r_0)Y_0(\lambda_n b)-J_0(\lambda_n b)Y_1(\lambda_n r_0)}.
\label{eq.Hr1}
\end{array}
\end{eqnarray}

It can be proved that the magnetic field solutions obtained in equations (\ref{eq.Hz1}) and (\ref{eq.Hr1}) are unique. The proof of the uniqueness theorem is attached in the Appendix.

\section{Numeral verification}
\label{sec3}
\begin{figure}
\center
\includegraphics[width=0.55\textwidth]{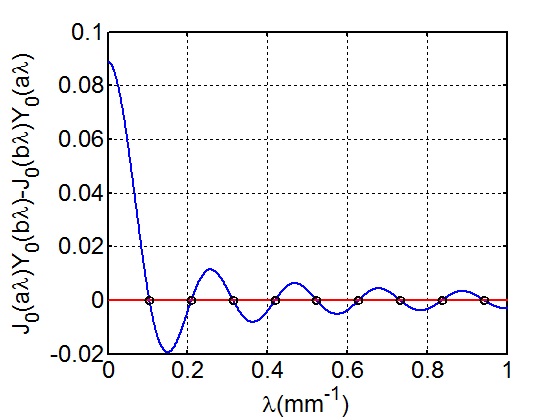}
\caption{The intersections of the function $J_0(\lambda a)Y_0(\lambda b)-J_0(\lambda b)Y_0(\lambda a)$ over $\lambda$ (the blue curve) and the horizontal axis (the red curve).}
\label{figure2}
\end{figure}
In order to evaluate the analytical algorithm accuracy, numerical simulations based on the finite element
method (FEM) are performed. In these FEM simulations, the parameters are set close to a 1:1 real watt balance magnet: $a$, $b$, $l$ and $r_0$ are set as $200$\,mm, $230$\,mm, $75$\,mm and $210$\,mm respectively; the relative permeability of the yoke is set as $\mu_r=100000$ and the magnetic strength of the permanent magnet is set as 800\,kAm$^{-1}$ in the vertical direction.

In the simulation, the magnetic flux density profile $B_r$(210\,mm, $z$) is calculated by FEM simulation as a known condition of the presented analytical algorithm to simulate the actual measurements of either a GC coil or a magnetic probe. Note that in reality it is impossible to calculate infinite number of terms $n$ in equation (\ref{eq.solution}), and hence the first nine $\lambda_n$ are adopted. The solution of the magnetic scalar potential $\varphi_m$, accordingly, can be written as
\begin{eqnarray}
\begin{array}{l}
\varphi_m=-r_0\left[\sum\limits_{i=0}^{9}F_i\mathbf{M}^{-1}(i,0)\right](\ln r-\ln b)\\
+\sum\limits_{n=1}^{9}\left[\sum\limits_{i=0}^{9}F_i\mathbf{M}^{-1}(i,n)\right]
\displaystyle\frac{\cosh (\lambda_n z)[J_0(\lambda_n r)Y_0(\lambda_n b)-J_0(\lambda_n b)Y_0(\lambda_n r)]}{\lambda_n\left[J_1(\lambda_n r_0)Y_0(\lambda_n b)-J_0(\lambda_n b)Y_1(\lambda_n r_0)\right]}.
\label{solution_4}
\end{array}
\end{eqnarray}

In equation (\ref{solution_4}), $\lambda_n(n=1,2,...,9)$, as demonstrated in solving equation (\ref{eq.condition12}), should satisfy $J_0(\lambda_n a)Y_0(\lambda_n b)-J_0(\lambda_n b)Y_0(\lambda_n a)=0$, and their values are calculated by a numerical method. As shown in figure \ref{figure2}, the solutions $\lambda_n$ are the intersections of function $J_0(\lambda a)Y_0(\lambda b)-J_0(\lambda b)Y_0(\lambda a)$ and the horizontal axis. The first nine intersections are solved as $\lambda_n=0.1047n~$mm$^{-1}$ ($n=1,2,...,9$).

As presented in equation (\ref{eq.orthogonal_2}), the base functions $\cosh(\lambda_n z)$ are transformed into orthogonal normalized functions $e_n (n=1,2,...,9)$, where $\lambda_0$ is set to $0$, as it is defined in section \ref{sec2}. Using equation (\ref{eq.orthogonal_31}), constants $F_i(i=1,2,...,9)$ are solved as
{\small\begin{equation}
F=\left(
\begin{array}{r}
5660653.35\\
-32140.89\\
-72396.98\\
243.16\\
3159.70\\
72.41\\
-189.91\\
5.03\\
10.43\\
-6.11\\
\end{array}
    \right).
\label{eq.numberal_1}
\end{equation}}
The $\mathbf M$ in this example is a $10\times10$ matrix, whose elements have been calculated based on equation (\ref{eq.matrix}), i.e.
\small\begin{equation}
    \mathbf{M}=\left(
      \begin{array}{llllllllll}
        1.2e1 & 0 & 0 & 0 & 0 & 0 & 0 & 0 & 0 &0\\
        2.0e3 & 3.4e3 & 0 & 0 & 0 & 0 & 0 & 0 & 0 &0\\
        2.6e6 & 6.4e6 & 2.2e6 & 0 & 0 & 0 & 0 & 0 & 0 &0\\
        4.4e9 & 1.3e10 & 7.0e9 & 1.4e9 & 0 & 0 & 0 & 0 & 0 &0\\
        8.6e12 & 2.7e13 & 1.8e13 & 6.4e12 & 9.1e11 & 0 & 0 & 0 & 0 &0\\
1.8e16	&5.7e16	&4.6e16	&2.1e16	&5.3e15	&5.9e14	&0	&0	&0	&0\\
3.8e19	&1.3e20	&1.1e20	&6.0e19	&2.1e19	&4.2e18	&3.8e17	&0	&0	&0\\
8.4e22	&2.9e23	&2.7e23	&1.6e23	&6.8e22	&1.9e22	&3.2e21	&2.5e20	&0	&0\\
1.9e26	&6.6e26	&6.5e26	&4.3e26	&2.1e26	&7.1e25	&1.7e25	&2.4e24	&1.6e23	&0\\
4.3e29	&1.5e30	&1.6e30	&1.1e30	&5.9e29	&2.4e29	&6.9e28	&1.4e28	&1.8e27	&1.0e26\\
      \end{array}
    \right).
\end{equation}
\normalsize
The matrix $\mathbf M^{-1}(i,n)$ is obtained by inversing $\mathbf M$, and two magnetic field components, i.e., $H_z(r,z)$ and $H_r(r,z)$, then can be solved based on equations (\ref{eq.Hz1}) and (\ref{eq.Hr1}).

\begin{figure}
\center
\includegraphics[width=0.55\textwidth]{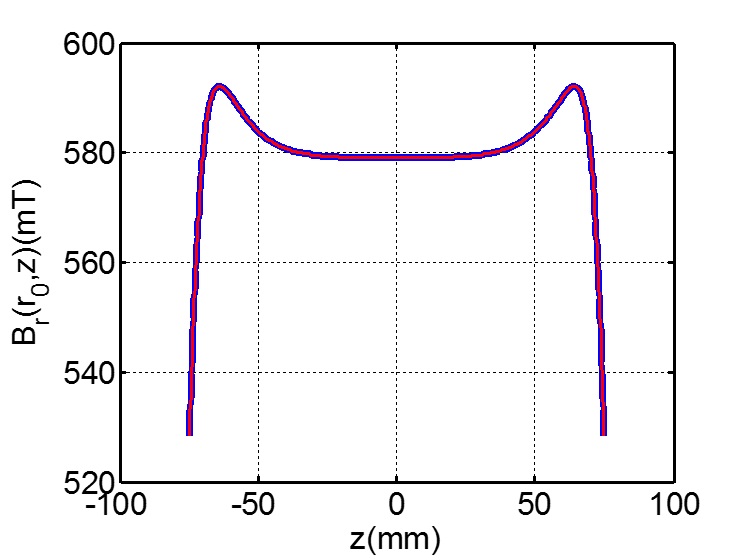}
\caption{The comparison between $B_r(r_0, z)$ obtained by FEM simulation (the blue curve) and the analytical method (the red curve).}
\label{figure25}
\end{figure}

\begin{figure}
\center
\includegraphics[width=0.55\textwidth]{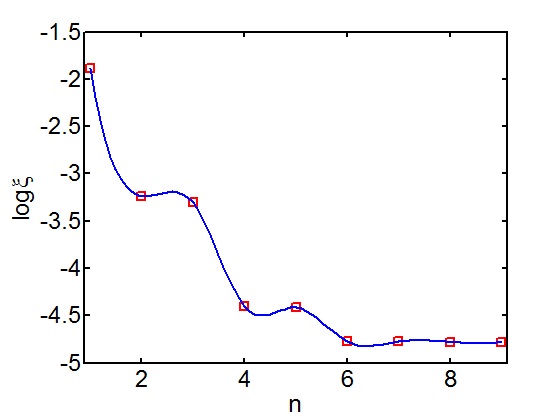}
\caption{The decay of the relative fit error $\xi$ over different orders $n$.}
\label{figure26}
\end{figure}

Figure \ref{figure25} shows the comparison of $B_r(r_0, z)$ results calculated by the FEM simulation and the analytical equation in (\ref{eq.Hr1}), where $B_r(r_0, z)=\mu_0H_r(r_0,z)$ and $\mu_0$ denotes the vacuum (air) permeability. It can be seen that the two curves agree well with each other, which indicates that the order of $n$ is high enough to estimate $B_r(r_0,z)$. In order to further check the agreement with different numbers of $n$, a relative fit error $\xi$ is defined, i.e.
\begin{eqnarray}
\begin{array}{l}
\displaystyle
\xi=\frac{1}{B_a}\sqrt{\frac{1}{N}\sum_{i=1}^{N}(B_i-B_{i0})^2},
\label{eq.constant}
\end{array}
\end{eqnarray}
where $B_{i0}$ is the measurement magnetic flux density, $B_i$ the fit value of the magnetic flux density, and $B_a$ the average of $B_i$, is analyzed on its decay rate over $n$. The calculation result is shown in figure \ref{figure26}. It can be seen that the fit accuracy is mainly limited by the magnetic field profile measurement. When $n=9$, the fit residual is already comparable to the sensitivity of the magnetic field measurement instruments, i.e. the accuracy of FEM simulation in this case.

\begin{figure}
\center
\includegraphics[width=0.9\textwidth]{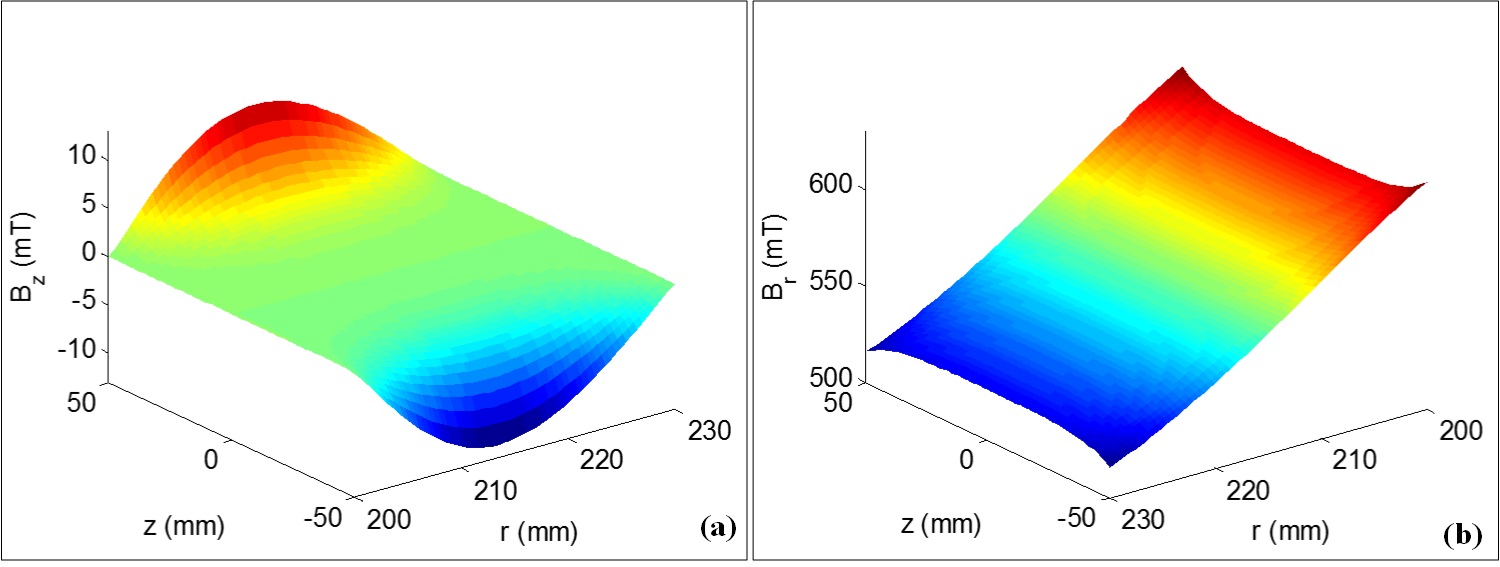}
\caption{Calculation results of magnetic flux density distribution. (a) Magnetic flux density distribution of $B_z(r,z)$. (b) Magnetic flux density distribution of $B_r(r,z)$. }
\label{figure4}
\end{figure}

Figure \ref{figure4} shows the calculation result of the 3D magnetic flux density $B_z(r, z)$ and $B_r(r, z)$ based on the analytical equations (\ref{eq.Hz1}) and (\ref{eq.Hr1}). The air gap region $r\in$(200\,mm, 230\,mm), $z\in$(-50\,mm, 50\,mm), where the watt balance is conventionally operated, has been focused. The calculation result clearly shows the fringe effect: The absolute value of the vertical magnetic flux density $B_z(r,z)$ increases at both vertical ends of the air gap; the horizontal magnetic flux density component $B_r(r,z)$ is also bent from the $1/r$ decay surface.

\begin{figure}
\center
\includegraphics[width=0.9\textwidth]{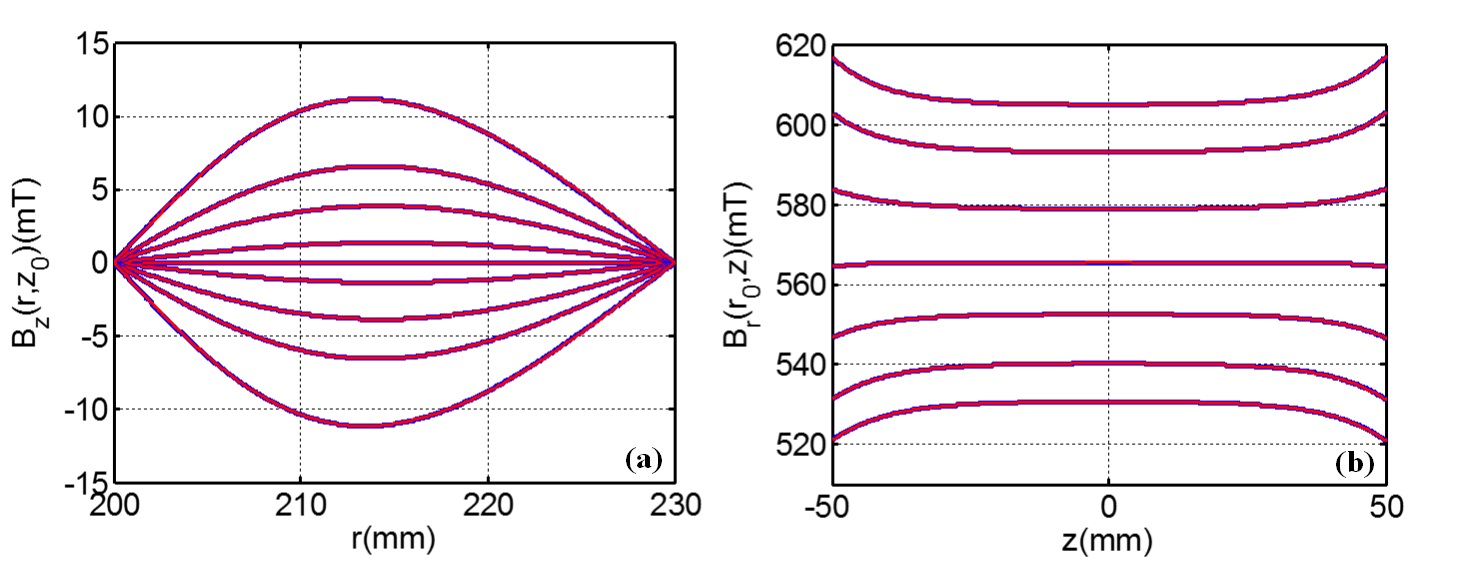}
\caption{Comparison of typical $B_z(r, z_0)$ and $B_r(r_0, z)$ curves obtained by the analytical method and FEM. The blue curves are obtained by FEM simulation and the red curves are calculated by the analytical method. (a) The $B_z(r, z_0)$ curves, which from top to bottom are presented with different values of $z_0$, i.e. 50\,mm, 45\,mm, 40\,mm, 30\,mm, 0\,mm, -30\,mm, -40\,mm, -45\,mm and -50\,mm. (b) The $B_r(r_0, z)$ curves, which from top to bottom have different values of $r_0$, i.e. 201\,mm, 205\,mm, 210\,mm, 215\,mm, 220\,mm, 225\,mm and 229\,mm.}
\label{figure3}
\end{figure}

\begin{figure}
\center
\includegraphics[width=0.95\textwidth]{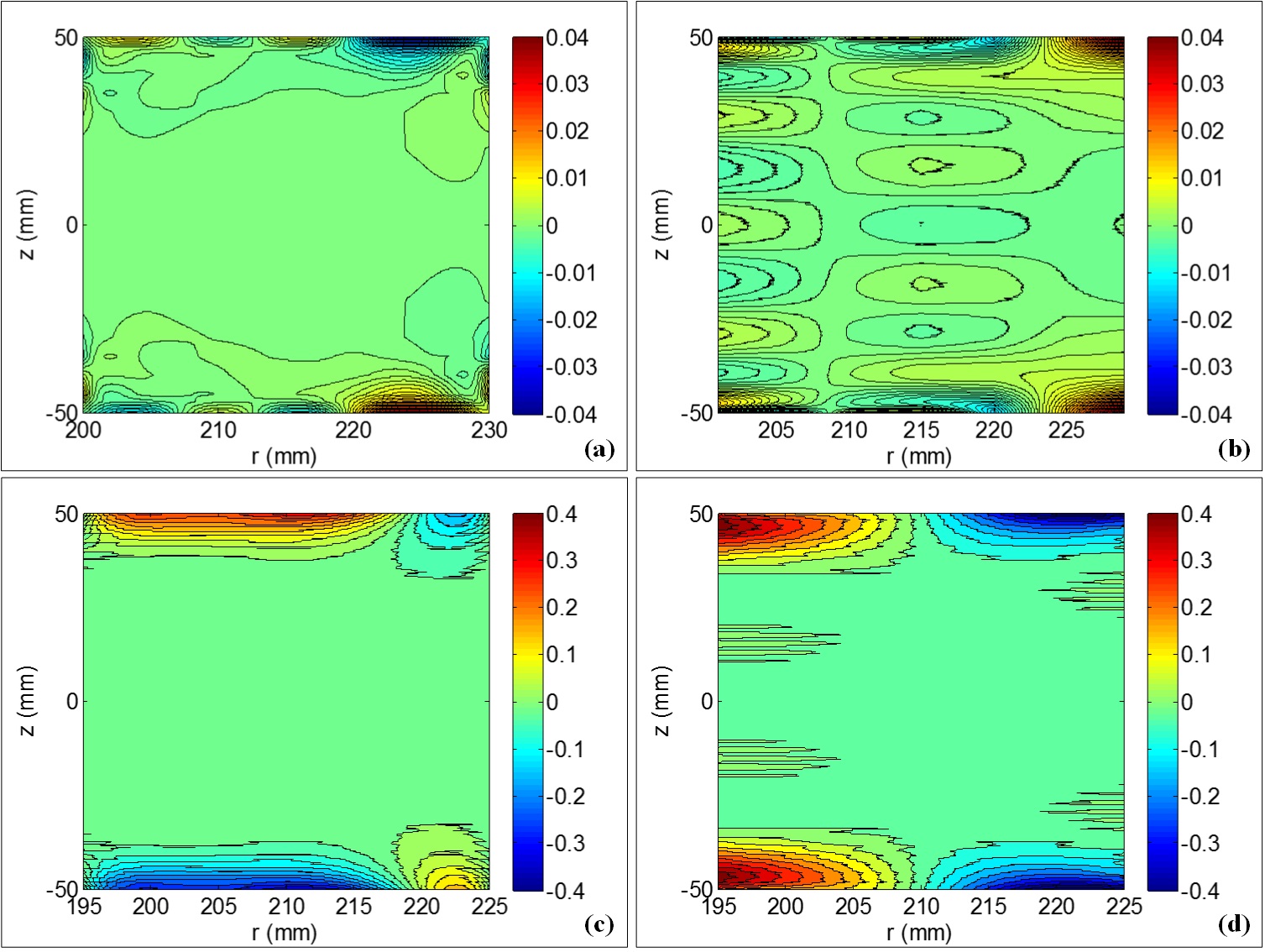}
\caption{The field calculation error in mT. The top subgraphs are the calculation error between the analytical algorithm and FEM simulation: (a) magnetic flux density error of $B_z(r, z)$; (b) the magnetic flux density error of $B_r(r, z)$. The bottom subgraphs are the calculation error between the polynomial estimation algorithm and FEM simulation: (c) magnetic flux density error of $B_z(r, z)$; (d) the magnetic flux density error of $B_r(r, z)$.}
\label{figure5}
\end{figure}

To evaluate the accuracy of the presented analytical algorithm, some typical $B_z(r, z_0)$ and $B_r(r_0, z)$ curves given by FEM simulation and the analytical equations (\ref{eq.Hz1}) and (\ref{eq.Hr1}) are compared in figure \ref{figure3}. It can be seen that the calculation obtained by the analytical algorithm agrees very well with that simulated by FEM. For a full view of the field difference of FEM and the analytical algorithm, the differential maps of $B_z(r,z)$ and $B_r(r,z)$ are calculated in the air gap region $r\in$(200\,mm, 230\,mm), $z\in$(-50\,mm, 50\,mm) and the calculation results are shown in figure \ref{figure5}.

For comparison of the two analytical algorithms, the field difference between the FEM simulation and the polynomial estimation algorithm in \cite{lishisong} has been also calculated with the same permeability. To quantize the comparison, the average difference between the analytical method and the FEM is defined as
\begin{equation}
\varsigma=\sqrt{\frac{1}{PQ}\sum_{i=1}^{P}\sum_{j=1}^{Q}\Delta B(i,j)^2},
\end{equation}
where $i=1,2,...,P$ is the radical index number and $j=1,2,...,Q$ is the vertical index number; $\Delta B$ is the magnetic flux density difference between the proposed analytical method and FEM simulation.
The polynomial estimation algorithm is adopted in its highest precision case, i.e. three different $B_r(z)$ profiles are measured in radii $201$\,mm, $205$\,mm and $210$\,mm. The average differences of the polynomial estimation algorithm are 0.0595\,mT for $B_z(r,z)$ and 0.0894\,mT for $B_r(r,z)$ while the average difference of the new analytical algorithm is 0.0060\,mT for $B_z(r,z)$ and 0.0048\,mT for $B_r(r,z)$. The accuracy of the new analytical algorithm in this case is respectively 10 and 20 times better for $B_z(r,z)$ and $B_r(r,z)$ than that of the polynomial estimation algorithm.

\section{Discussion}
\label{sec4}
\begin{figure}
\center
\includegraphics[width=0.95\textwidth]{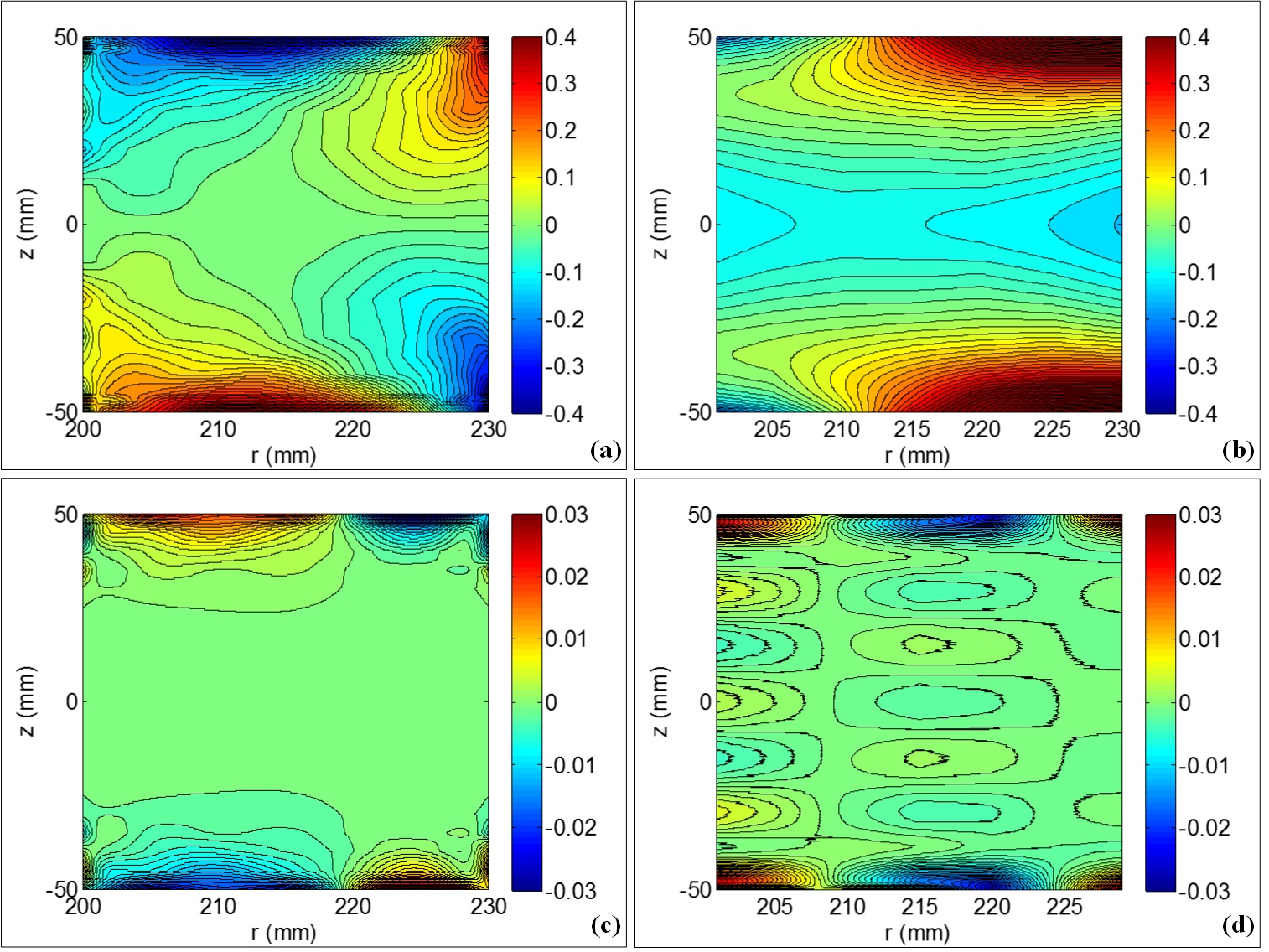}
\caption{The field calculation error in mT between the analytical algorithm and FEM simulation. The top subgraphs are the calculation error map when the relative permeability is set as 1000: (a) the calculation error of $B_z(r, z)$; (b) the calculation error of $B_r(r, z)$. The bottom subgraphs are the calculation error map when the relative permeability is set as $10^{14}$: (c) the calculation error of $B_z(r, z)$; (d) the calculation error of $B_r(r, z)$.}
\label{fig9}
\end{figure}

The numerical simulation in section \ref{sec3} exhibits advantages in field representation accuracy and convenience in measurement for the new analytical algorithm.
It should be emphasized that the shown analytical algorithm, as well as the polynomial estimation algorithm proposed in \cite{lishisong}, is based on the assumption that the normal component of the magnetic field on the air-yoke boundary strictly equals zero. However, in reality a weak normal component exists on the air-yoke boundaries duo to the finite permeability of the yoke, which creates the main part of the calculation error. For a further check, we calculated the magnetic field difference with different values of the yoke permeability. Figure \ref{fig9} shows the calculation errors between the proposed analytical algorithm and FEM simulation when $\mu_r=1000$ and $\mu_r=10^{14}$. It can be seen that the difference error is getting smaller when the permeability is higher. This indicates that the analytical algorithm will have better performance when applied in the watt balance magnet with high permeability, e.g. the BIPM watt balance magnet.

\begin{figure}
\center
\includegraphics[width=0.95\textwidth]{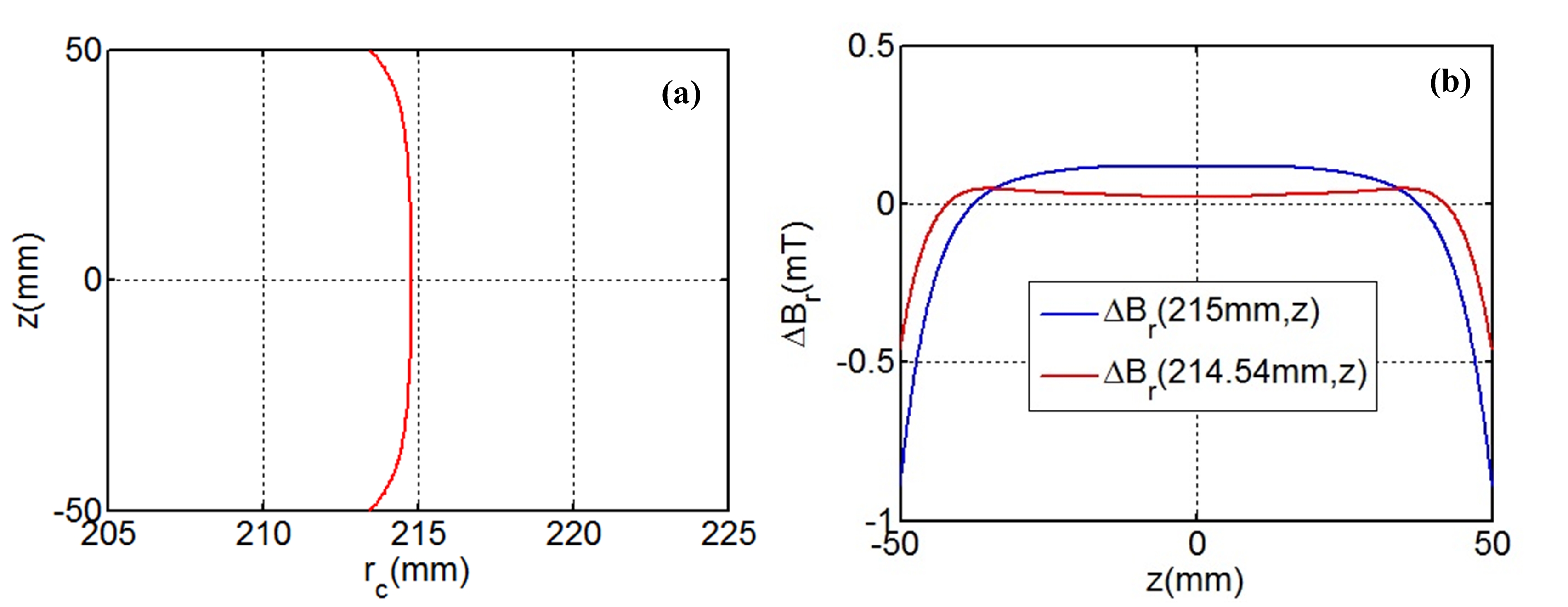}
\caption{(a) The calculation result of $r_c(z)$. (b) The simulation results of $B_r(214.54\,$mm$,z)$ and $B_r(215\,$mm$,z)$. $\Delta B_r$ is defined as $\Delta B_r=B_r(z)-B_{ra}$ where $B_{ra}$ is the average value of $B_r(z)$.}
\label{figure10}
\end{figure}

As mentioned in section \ref{sec1}, one of the main purpose for presenting the 3D magnetic field in the air gap is to find a best coil diameter for obtaining a widest $B_r(z)$ profile. Here we give a suggestion by taking the calculation in section \ref{sec3} as an example. Knowing the 3D magnetic field, the coordinate $r_c(z)$ where $B_z(r_c,z)$ has a peak value, i.e.
\begin{equation}
\begin{array}{l}
\displaystyle \left.\frac{\partial H_z(r,z)}{\partial r}\right|_{r=r_{c}}=\sum\limits_{n=1}^{\infty}\left[\sum\limits_{i=0}^{\infty}F_i\mathbf{M}^{-1}(i,n)\right]\\
\displaystyle\times\frac{\lambda_n\sinh (\lambda_n z_0)\left[J_1(\lambda_n r_{\rm p})Y_0(\lambda_n b)-J_0(\lambda_n b)Y_1(\lambda_n r_{\rm p})\right]}{J_1(\lambda_n r_0)Y_0(\lambda_n b)-J_0(\lambda_n b)Y_1(\lambda_n r_0)}=0,
\label{new}
\end{array}
\end{equation}
can be solved.
The calculation result of $r_c(z)$ has been shown in figure \ref{figure10}(a). It shows that for the typical watt balance magnet, a best coil radius should be smaller than the air gap center radius. If the velocity sweep range is set $(z_1, z_2)$ in the velocity mode, the best coil radius is calculated by means of the $r_c$ as
\begin{equation}
r_{coil}=\frac{\displaystyle\int_{z_1}^{z_2}r_c(z)\mbox{d}z}{z_2-z_1}.
\end{equation}
For the example shown in figure \ref{figure10}(a), $z_1=-50$\,mm, $z_2=50$\,mm, and $r_{coil}$ is calculated as 214.54\,mm. By the calculation, the $B_r(214.54\,$mm$,z)$ should have a wider flat profile than $B_r(215\,$mm$,z)$. To check it, we have plotted these two $B_r(z)$ profiled in figure \ref{figure10}(b) by FEM and the result verifies the conclusion.

\section{Conclusion}
\label{sec5}
An analytical algorithm, employing only one measured $B_r(z)$, is presented for calculating the 3D magnetic field profile in watt balance magnets. Compared to the polynomial estimation algorithm \cite{lishisong}, the new algorithm is based on fundamental electromagnetic natures of the magnet and has significant advantages in both convenience and accuracy. It is shown that the new analytical algorithm can improve the field mapping accuracy by more than one magnitude than that of a simple polynomial estimation, which has a good potential in application of high permeability cases, e.g., the BIPM magnet.

The presented work can supply necessary information for misalignment analysis and parameter determinations in watt balances. Base on the study, a best coil radius, which should be designed slightly smaller than the air gap center, is suggested. The discussion shows that the accuracy of the proposed analytical algorithm is mainly limited by a finite permeability of the yoke material. Therefore, a correction model of yoke permeability should be focused in a following investigation. Also, in this paper the $r$ symmetry of the watt balance magnet is assumed, which in reality may be not true. Further studies with considerations of the $r$ asymmetry of the magnet system may be addressed in the future.

\section*{Acknowledgement}
The authors would like to thank Mr. Xuanyu Dong for advice about using the Gram-Schmidt Orthogonalization Procedure and Dr Qing Wang at Durham University, UK for language proofing. This work is supported by the National Natural Science Foundation of China (Grant No. 51507088).

\section*{Appendix}
In this appendix, the uniqueness of magnetic field solutions, i.e., $H_r(r,z)$ and $H_z(r,z)$ in equations (\ref{eq.Hz1}) and (\ref{eq.Hr1}), is proved by a reductio ad absurdum.
We assume that the solution of equation (\ref{eq.solution}) is not unique in the magnet air gap region $r\in(a,b)$, $z\in(-l,l)$. Without losing generality, two different solutions $\varphi_{m1}(r,z)$ and $\varphi_{m2}(r,z)$ are supposed to satisfy the condition $H_z(a,z)=0$, $H_z(b,z)=0$ and $H_r(r_0,z)$. Based on the symmetry of the air gap, the magnetic scalar potential at $z=\pm l$ for $\varphi_{m1}(r,z)$ and $\varphi_{m2}(r,z)$, i.e. $\varphi_{m1}(r,l)$, $\varphi_{m1}(r,-l)$, $\varphi_{m2}(r,l)$ and $\varphi_{m2}(r,-l)$, should meet
\begin{equation}
\begin{array}{l}
\varphi_{m1}(r,l)=\varphi_{m1}(r,-l),\\
\varphi_{m2}(r,l)=\varphi_{m2}(r,-l).
\end{array}
\end{equation}
According to the uniqueness theorem for static magnetic fields, $H_z(a,z)=0$, $H_z(b,z)=0$, $\varphi_{m1}(r,l)$ and $\varphi_{m1}(r,-l)$ form the boundary conditions of region $r\in(a,b)$, $z\in(-l,l)$ and determines a unique solution of $\varphi_{m1}(r,z)$, which is expressed similarly as equation (\ref{eq.after_condition2}), i.e.
\begin{eqnarray}
\begin{array}{l}
\varphi_{m1}(r,z)=C(\ln r-\ln b)\\
+\sum\limits_{n=1}^{\infty}A_{1n}\cosh (\lambda_n z)
\left[J_0(\lambda_n r)Y_0(\lambda_n b)-J_0(\lambda_n b)Y_0(\lambda_n r)\right].
\end{array}
\label{eq.app1}
\end{eqnarray}
The unknown constants in equation (\ref{eq.app1}), i.e. $C$ and $A_{1n}$, can be solved by expanding $\varphi_{m1}(r,l)$ in forms of fundamental functions $\ln r-\ln b$ and $J_0(\lambda_n r)Y_0(\lambda_n b)-J_0(\lambda_n b)Y_0(\lambda_n r)$. The solved constants are symbolized as $C^{m1}$ and $A_{1n}^{m1}$, and then the solution $\varphi_{m1}(r,z)$ can be written as
\begin{eqnarray}
\begin{array}{l}
\varphi_{m1}(r,z)=C^{m1}(\ln r-\ln b)\\
+\sum\limits_{n=1}^{\infty}A_{1n}^{m1}\cosh (\lambda_n z)
\left[J_0(\lambda_n r)Y_0(\lambda_n b)-J_0(\lambda_n b)Y_0(\lambda_n r)\right].
\label{eq.phim1}
\end{array}
\end{eqnarray}

Similarly, $\varphi_{m2}(r,z)$ is the unique solution with boundary conditions $H_z(a,z)=0$, $H_z(b,z)=0$, $\varphi_{m2}(r,l)$ and $\varphi_{m2}(r,-l)$, which can be expressed as
\begin{eqnarray}
\begin{array}{l}
\varphi_{m2}(r,z)=C^{m2}(\ln r-\ln b)\\
+\sum\limits_{n=1}^{\infty}A_{1n}^{m2}\cosh (\lambda_n z)
\left[J_0(\lambda_n r)Y_0(\lambda_n b)-J_0(\lambda_n b)Y_0(\lambda_n r)\right].
\label{eq.phim2}
\end{array}
\end{eqnarray}
Based on equations (\ref{eq1}), (\ref{eq1.1}) and (\ref{eq.phim1}), the radial component of the magnetic field at $r_0$, i.e. $H_r^{m1}(r_0,z)$,
can be expressed as
\begin{eqnarray}
\begin{array}{l}
H_r^{m1}(r_0,z)=\displaystyle\left.-\frac{\partial \varphi_{m1}(r,z)}{\partial r}\right|_{r=r_0}\\
\displaystyle=-\frac{C^{m1}}{r_0}
+\sum\limits_{n=1}^{\infty}A_{1n}^{m1}\lambda_n
\left[J_1(\lambda_n r_0)Y_0(\lambda_n b)-J_0(\lambda_n b)Y_1(\lambda_n r_0)\right]\cosh (\lambda_n z),~~~~
\end{array}
\end{eqnarray}
and based on equations (\ref{eq1}), (\ref{eq1.1}) and (\ref{eq.phim2}), the expression of $H_r^{m2}(r_0,z)$ is obtained as
\begin{eqnarray}
\begin{array}{l}
H_r^{m2}(r_0,z)=\displaystyle\left.-\frac{\partial \varphi_{m2}(r,z)}{\partial r}\right|_{r=r_0}\\
\displaystyle=-\frac{C^{m2}}{r_0}
+\sum\limits_{n=1}^{\infty}A_{1n}^{m2}\lambda_n
\left[J_1(\lambda_n r_0)Y_0(\lambda_n b)-J_0(\lambda_n b)Y_1(\lambda_n r_0)\right]\cosh (\lambda_n z).~~~~
\end{array}
\end{eqnarray}

It has been supposed that both $\varphi_{m1}(r,z)$ and $\varphi_{m2}(r,z)$ can both satisfy the condition $H_r(r_0,z)$, therefore $H_r^{m1}(r_0,z)=H_r^{m2}(r_0,z)$. As is known, a function of $\cosh(\lambda_m z)$ cannot be expressed by linear combinations of $\cosh(\lambda_n z)$ where $n\neq m$, thus $C_{m1}=C_{m2}$ and $A_{1n}^{m1}=A_{1n}^{m2}$ are established. Further, based on equations (\ref{eq.phim1}) and (\ref{eq.phim2}), we have
\begin{equation}
\varphi_{m1}(r,z)=\varphi_{m2}(r,z).
\label{eq.appxZ}
\end{equation}
Obviously, equation (\ref{eq.appxZ}) and the stated assumption $\varphi_{m1}(r,z)\neq\varphi_{m2}(r,z)$ are contradictory, and hence the supposition is false and the theorem, i.e. the solution of equation (\ref{eq.solution}) is unique, is valid. Therefore $H_r(r,z)$ and $H_z(r,z)$ in equations (\ref{eq.Hz1}) and (\ref{eq.Hr1}) are the unique solutions.




\begin{thebibliography}{0}%
\makeatletter
\providecommand \@ifxundefined [1]{%
 \@ifx{#1\undefined}
}%
\providecommand \@ifnum [1]{%
 \ifnum #1\expandafter \@firstoftwo
 \else \expandafter \@secondoftwo
 \fi
}%
\providecommand \@ifx [1]{%
 \ifx #1\expandafter \@firstoftwo
 \else \expandafter \@secondoftwo
 \fi
}%
\providecommand \natexlab [1]{#1}%
\providecommand \enquote  [1]{``#1''}%
\providecommand \bibnamefont  [1]{#1}%
\providecommand \bibfnamefont [1]{#1}%
\providecommand \citenamefont [1]{#1}%
\providecommand \href@noop [0]{\@secondoftwo}%
\providecommand \href [0]{\begingroup \@sanitize@url \@href}%
\providecommand \@href[1]{\@@startlink{#1}\@@href}%
\providecommand \@@href[1]{\endgroup#1\@@endlink}%
\providecommand \@sanitize@url [0]{\catcode `\\12\catcode `\$12\catcode
  `\&12\catcode `\#12\catcode `\^12\catcode `\_12\catcode `\%12\relax}%
\providecommand \@@startlink[1]{}%
\providecommand \@@endlink[0]{}%
\providecommand \url  [0]{\begingroup\@sanitize@url \@url }%
\providecommand \@url [1]{\endgroup\@href {#1}{\urlprefix }}%
\providecommand \urlprefix  [0]{URL }%
\providecommand \Eprint [0]{\href }%
\providecommand \doibase [0]{http://dx.doi.org/}%
\providecommand \selectlanguage [0]{\@gobble}%
\providecommand \bibinfo  [0]{\@secondoftwo}%
\providecommand \bibfield  [0]{\@secondoftwo}%
\providecommand \translation [1]{[#1]}%
\providecommand \BibitemOpen [0]{}%
\providecommand \bibitemStop [0]{}%
\providecommand \bibitemNoStop [0]{.\EOS\space}%
\providecommand \EOS [0]{\spacefactor3000\relax}%
\providecommand \BibitemShut  [1]{\csname bibitem#1\endcsname}%
\let\auto@bib@innerbib\@empty
\end{thebibliography}%


\begin{thebibliography}{1}

\bibitem{lishisong}
Li S {\it et al} 2015 Field representation of a watt balance magnet by partial profile measurements {\it Metrologia} {\color{blue}{\bf52} 445-453}

\bibitem{Kibble1976}
Kibble B P 1976 A measurement of the gyromagnetic ratio of the proton by the strong field method {\it Atomic Masses and Fundamental Constants 5} {\color{blue}pp 545-551}

\bibitem{ShisongLi2012}
Robinson I 2009 Toward the redefinition of the kilogram: measurements of planck's constant using watt balances {\it IEEE Trans. Instrum. Meas.}
{\color{blue}{\bf51} 942-948}

\bibitem{new_Planck}
Mills I {\it et al} 2006 Redefinition of the kilogram, ampere, kelvin and mole: a proposed approach to implementing CIPM recommendation 1 (CI-2005) {\it Metrologia}
{\color{blue}{\bf43} 227-246}

\bibitem{Li12}
Li S {\it et al} 2012 Precisely measuring the Planck constant by electromechanical balances {\it Measurement} {\color{blue}{\bf 45} 1-13}

\bibitem{Stainer13}
Steiner R 2013 History and progress on accurate measurements of the Planck constant {\it Reports on Progress in Physics} {\color{blue}{\bf 76} 016101}

\bibitem{Stock13}
Stock M 2013 Watt balance experiments for the determination of the Planck constant and the redefinition of the kilogram {\it Metrologia} {\color{blue}{\bf 50} R1-R16}

\bibitem{NPL}
Robinson I 2012 Towards the redefinition of the kilogram: a measurement of the Planck constant using the NPL Mark II watt balance {\it Metrologia}
{\color{blue}{\bf49} 113-156}

\bibitem{NIST}
Schlamminger S {\it et al} 2014 Determination of the Planck constant using a watt balance with a superconducting magnet system at the National Institute of Standards and Technology {\it Metrologia}
{\color{blue}{\bf51} S15-S24}

\bibitem{METAS}
Eichenberger A {\it et al} 2011 Determination of the Planck constant with the METAS watt balance {\it Metrologia}
{\color{blue}{\bf48} 133-141}

\bibitem{BIPM}
Picard A {\it et al} 2011 The BIPM watt balance: improvements and developments {\it IEEE Trans. Instrum. Meas.}
{\color{blue}{\bf60} 2378-2386}

\bibitem{LNE}
Genev\`{e}s G {\it et al} 2005 The BNM watt balance project {\it IEEE Trans. Instrum. Meas.}
{\color{blue}{\bf54} 850-853}

\bibitem{NIM}
Zhang Z {\it et al} 2014 The joule balance in NIM of China {\it Metrologia} {\color{blue}{\bf51} S25-S31}

\bibitem{NRC}
Sanchez A {\it et al} 2014 A determination of Planck's constant using the NRC watt balance {\it Metrologia}
{\color{blue}{\bf51} S5-S14}

\bibitem{MSL}
Sutton C 2009 An oscillatory dynamic mode for a watt balance {\it Metrologia}
{\color{blue}{\bf46} 467-472}

\bibitem{KRISS}
Kim D {\it et al} 2014 Design of the KRISS watt balance {\it Metrologia}
{\color{blue}{\bf51} S96-S100}

\bibitem{Stock2006}
Stock M 2006 Watt balances and the future of the kilogram {\it INFOSIM Informative Bulletin of the Inter American Metrology System (November 2006)-OAS}
{\color{blue}pp. 9-13}

\bibitem{lishisongcoil}
Li S {\it et al} 2015 Coil motion effects in watt balances: a theoretical check

\bibitem{Baumann2013}
Baumann H {\it et al} 2013 Design of the new METAS watt balance experiment Mark II {\it Metrologia}
{\color{blue}{\bf50} 235-242}

\bibitem{Seifert2014}
Seifert F {\it et al} 2014 Construction, Measurement, Shimming, and Performance of the NIST-4 Magnet System {\it IEEE Trans. Instrum. Meas.}
{\color{blue}{\bf63} 3027-3038}

\bibitem{Sutton2014}
Sutton C {\it et al} 2014 A magnet system for the MSL watt balance {\it Metrologia}
{\color{blue}{\bf51} S101-S106}

\bibitem{Robinson}
Robinson I 2012 Alignment of the NPL Mark II watt balance {\it Meas. Sci. Technol.}
{\color{blue}{\bf23} 124012}

\end{thebibliography}
\end{document}